\font\mybb=msbm10 at 12pt
\def\bbbb#1{\hbox{\mybb#1}}
\def\Z {\bbbb{Z}}
\def\R {\bbbb{R}}

\def \aa {\alpha}
\def \bb {\beta}
\def \gg {\gamma}
\def \dd {\delta}
\def \ee {\epsilon}

\def \mm {\mu}
\def \nn {\nu}

\def \rr {\rho}
\def \ss {\sigma}

\def \th {\theta}

\def \ww{\omega}

\def \fff {\Phi}
\def \lll {\wedge}

\def \ti {\tilde}

\def \2 {{1 \over 2}}
\def \3 {{1 \over 3}}
\def \4 {{1 \over 4}}
\def \5 {{1 \over 5}}
\def \6 {{1 \over 6}}
\def \7 {{1 \over 7}}
\def \8 {{1 \over 8}}
\def \9 {{1 \over 9}}
\def \0 { \infty}

\def\++ {{(+)}}
\def \- {{(-)}}
\def\+-{{(\pm)}}

\def \pa {\partial}


\tolerance=10000
\input phyzzx

 \def\unit{\hbox to 3.3pt{\hskip1.3pt \vrule height 7pt width .4pt \hskip.7pt
\vrule height 7.85pt width .4pt \kern-2.4pt
\hrulefill \kern-3pt
\raise 4pt\hbox{\char'40}}}

\def\nup#1({Nucl.\ Phys.\  {\bf B#1}\ (}

\def\tr{{\rm tr}}
\def\Str{{\rm Str}}
\REF\Mat{T. Banks, W. Fischler, S. Shenker, and L. Susskind,
``M theory as a matrix model: a conjecture,''
hep-th/9610043, Phys. Rev.
{\bf D55} (1997) 5112.}%
\REF\brev{For a review see T. Banks,
hep-th/9706168, hep-th/9710231.}%
\REF\Sus{L. Susskind, hep-th/9704080.}
\REF\taylor{W. Taylor, hep-th/9611042, Phys. Lett. {\bf B394} (1997) 283.}%
\REF\motl{L. Motl,hep-th/9701025.}
\REF\IIAbs{T. Banks and N. Seiberg,hep-th/9702187.}
\REF\IIAVV{R. Dijkgraaf,E. Verlinde and H. Verlinde, hep-th/9703030.}
\REF\dvv{R. Dijkgraaf, E. Verlinde, and H. Verlinde,  hep-th/9603126;
 hep-th/9604055; hep-th/9703030; hep-th/9704018.}
\REF\rozali{M. Rozali,hep-th/9702136.}
\REF\brs{M. Berkooz, M. Rozali, and N. Seiberg,
hep-th/9704089}
\REF\fhrs{W. Fischler, E. Halyo, A. Rajaraman and L. Susskind,
hep-th/9703102.}%
\REF\MatTor{N. Seiberg,
hep-th/9705221.}%
\REF\Hv{F. Hacquebord and H. Verlinde, hep-th/9707179.}
\REF\Alg{S. Elitzur, A. Giveon, D. Kutasov and E. Rabinbovici, hep-th/9707217.}
\REF\mem{A. Losev, G. Moore and S.L. Shatashvili, hep-th/9707250; I. Brunner
and A. Karch,
hep-th/9707259; A. Hanany and G. Lifschytz, hep-th/9708037. }
\REF\Sen{A. Sen, hep-th/9709220.}
\REF\Seib{N. Seiberg,
hep-th/9710009.}%
\REF\prog{C.M. Hull, in preparation}
\REF\progj{C.M. Hull and B. Julia, in preparation}
\REF\Dis {D. Berenstein, R. Corrado and J. Distler, hep-th/9704087.}
\REF\HT{C.M. Hull and P.K. Townsend, hep-th/9410167.}
\REF\town{P.K. Townsend, hep-th/9501068.}
\REF\GravDu{C.M. Hull, hep-th/9705162.}
\REF\bergy{E. Bergshoeff, M. de Roo, M.B. Green, G. Papadopoulos and P.K.
Townsend, Nucl. phys. {\bf B470} (1996) 113.}
\REF\Ym{C.M. Hull, hep-th/9710165}
\REF\Dbrane{R.G. Leigh, Mod. Phys. Lett. {\bf A28 }  (1989) 2767;
 C. Schmidhuber, hep-th/9601003; M.B. Green, C.M. Hull
and P.K.
Townsend, Phys. Lett. {\bf B382} (1996) 65.}
\REF\Doug{M. Douglas, hep-th/9512077.}
\REF\polch{S. Chaudhuri, C. Johnson, and J. Polchinski,
``Notes on D-branes,'' hep-th/9602052; J. Polchinski,
``TASI Lectures on D-branes,'' hep-th/9611050}
\REF\Cal{C.G. Callan, C. Lovelace, C.R. Nappi and S.A. Yost, Nucl. Phys. {\bf
B308} (1988)
221.}
\REF\PL{H. Lu and C.N. Pope, hep-th/9701177}
\REF\tonn{P. K. Townsend, hep-th/9708034.}
\REF\mar{E. Martinec, hep-th/9706194.}



\Pubnum{ \vbox{  \hbox {QMW-97-35} \hbox{LPTENS 97/57}\hbox{hep-th/9711179}} }
\pubtype{}
\date{November, 1997}

\titlepage

\title {\bf  Matrix Theory, U-Duality and Toroidal Compactifications of
M-Theory }

\author{C.M. Hull}
\address{Physics Department, Queen Mary and Westfield College,
\break Mile End Road, London E1 4NS, U.K.}
\andaddress{Laboratoire de Physique Th\' eorique, Ecole Normale Sup\' erieure,
24 Rue Lhomond, 75231 Paris Cedex 05, France.}


\abstract {
Using U-duality, the properties of the
matrix theories  corresponding to the compactification of M-theory  on $T^d$
are investigated.
The couplings of the $d+1$ dimensional effective Super-Yang-Mills theory  to
all the M-theory moduli
is deduced
and the spectrum of BPS branes in the SYM gives the corresponding spectrum of
the matrix theory.
Known results are recovered for $d\le 5$ and   predictions for $d>5$ are
proposed.
For $d>3$, the spectrum includes $d-4$ branes arising from
YM instantons, and   U-duality interchanges momentum modes with brane wrapping
modes.
For $d=6$, there is a generalised $\th $-angle which couples to instantonic
3-branes
and which combines with the SYM coupling constant to take values in
$SL(2,\R)/U(1)$, acted on by an
$SL(2,\Z)$ subgroup of the U-duality group $E_6(\Z)$.
For $d=4,7,8$, there is an $SL(d+1)$ symmetry, suggesting that the matrix
theory could be
a scale-invariant $d+2$ dimensional theory on $T^{d+1} \times \R$ in these
cases, as is already
known to be the case for   $d=4$;   evidence is found suggesting this happens
for $d=8$ but not
$d=7$. }

\endpage

\chapter {Matrices and Branes}

M-theory in the infinite momentum frame is conjectured to be described by the
large $N$ limit of the
$U(N)$ matrix quantum mechanics obtained from reducing 10-D super Yang-Mills to
one dimension
[\Mat,\brev]. M-theory compactified on $T^d$ for $d\le 3$ is conjectured to be
described by a $d+1$
dimensional matrix theory given by
$d+1$ dimensional Super-Yang-Mills (SYM) on $\tilde T^d\times \R$ where $\tilde
T^d$ is the dual
torus [\Mat-\Seib]. For $d>3$, the matrix description is given approximately at
low energies by $d+1$
dimensional  SYM,  but as the theory is non-renormalizable, extra degrees of
freedom must become important at short distances. These should give rise to a
consistent quantum
matter theory (i.e. one without gravity) and this has been shown to be the case
for $d \le 5$.
For $d=4$, this is the $5+1$ dimensional (2,0) supersymmetric self-dual
tensor multiplets with $U(N)$ gauge symmetry [\rozali-\MatTor], and for $d=5$
it is the (2,0)
supersymmetric non-critical  self-dual string theory [\dvv-\MatTor].  In each
of the  cases with $d
\le 5$,
 the U-duality group is manifest in the matrix description. U-duality then
defines much of the
structure of the matrix theory, and we will here investigate theories which
have an effective
description in terms of SYM on
$\tilde T^d\times \R$ and which are invariant under the appropriate U-duality
group. Recently a
matrix theory for $d=6$ has been discussed [\mem], and some difficulties
pointed out [\Sen,\Seib].
In [\Sus] it was conjectured that M-theory on   $T^{d+1}$, after an
 infinite boost so that one of the
circles becomes null, should be  described, in a suitable limit, by a similar
matrix theory whose
low-energy limit is
$d+1$ dimensional SYM on $\ti T^d \times \R$ with $U(N)$ gauge group, for {\it
finite } $N$.
A derivation of this, and of  the conjecture of [\Mat],
 has been proposed in [\Sen,\Seib].

If a matrix theory exists for
 $d \ge 6$, then it presumably requires the existence of some \lq exotic'
matter theory with $U(N)$
gauge symmetry  which (i) reduces to $d+1$ dimensional  SYM at low energies
(ii) reduces to a
non-critical string theory in the limit in which $T^d$ decompactifies to
$T^5\times
\R^{d-5}$, so that it should contain extended objects (iii) has \lq manifest'
U-duality under the
U-duality group $E_d(\Z)$ (where $E_5=SO(5,5)$, $E_4=SL(5)$,
$E_3=SL(3)\times SL(2)$ and  $E_2=SL(2)\times \R$). The conjecture of [\Sus]
suggests that for finite
$N$, in addition the SYM on $T^{d}\times \R$ should satisfy the condition (iv)
that it should be
invariant under
$E_{d+1}(\Z)$, the duality group for compactification on $T^{d+1}$.
Indeed, in [\Hv] it was suggested that for 4-dimensional SYM on $T^3\times \R$
the
expected $SL(3,\Z)\times SL(2,\Z)$ duality symmetry
 (the $SL(3,\Z)$ symmetry of $T^3$ and the S-duality of 3+1 dimensional SYM)
could be extended to $SL(5,\Z)$ by
Nahm-type transformations mixing the rank $N$ with
electric and magnetic fluxes. It was shown that the BPS spectrum indeed fits
into $SL(5,\Z)$
representations. In [\prog], evidence for this will be presented for $d>3$; the
analysis of [\Alg]
will be generalised to show that a large class of BPS states, shown to fit into
representations of
$E_{d}(Z)$ in [\Alg], in fact  fit into representations of $E_{d+1}(Z)$.
In [\progj], it will be shown that the U-duality group for toroidal
compactifications of M-theory
are unchanged in the limit in which one of the circles becomes null.
The properties of the effective SYM theory give a great deal of information
about the full
matrix theory, much in the same way that
the effective supergravity theories give information about non-perturbative
string theories [\HT],
and analysing  these is the aim of this paper. In particular, BPS solitonic
branes of the SYM theory
imply the existence of BPS states of the matrix theory which can be
extrapolated to BPS states in   regimes where the SYM is not a good effective
description,
  just as solitonic branes of supergravity
correspond to fundamental strings, D-branes and solitonic branes of
non-perturbative string theory
[\HT].

The same $U(N)$ SYM theories in $d+1$ dimensions also emerge in the effective
description of $N$
coincident Dd-branes (i.e. D-branes with $d+1$ dimensional world-volume). This
is of course not an
accident; the matrix theory emerges from the description of $N$ D-branes in a
limit in
which (at least for low enough $d$) the bulk fields  and the massive string
modes decouple
[\Mat,\Sen,\Seib]. The uncompactified M-theory is formulated in terms of
$N$ 0-branes in the limit $N \rightarrow \infty$ [\Mat] and this is also a
reasonable approximate description
for M-theory on $T^d$ if the torus is large. However, this description is
missing the string winding
modes which become light if the torus is small. For a small torus, T-duality
can be used to
transform $N$ 0-branes on $T^d$ to $N$ Dd-branes wrapped around the dual torus
$\ti T^d$. The
perturbative dynamics of the D-branes is given in terms of fundamental strings
ending on the
D-branes, and the low-energy effective action arising from this for $N$ Dd
branes is $d+1$
dimensional $U(N)$ SYM. For $d \le 3$ the SYM is a well-defined quantum theory
and provides a good
description of the D-brane dynamics at energies much less than the string
scale. For $d \ge 4$,
however, SYM must be supplemented by other degrees of freedom. For example, the
D4-brane of the IIA
theory  at finite string coupling
is given by the M5-brane wrapped around the circular dimension
of M-theory compactified on $S^1$.    The dynamics are then given in terms of
the effective
world-volume theory of the M5-brane, which is the 5+1 dimensional (2,0)
supersymmetric self-dual
tensor theory.

In each case, for low enough  $d$,
there is a limit in which the string length becomes infinite, supergravity  and
the
tower of massive string states decouple leaving a matter theory with a complete
quantum
description.
As  Dd-branes with different values of $d$ are all related by T-duality,
it is natural to expect  such a limit  for all values of $d$, so that there
should be
a consistent quantum matter theory for
$d \ge 5$ that reduces to $U(N)$ SYM at  low energy. By T-duality, it would be
this limiting theory,
in the additional limit $N\rightarrow \infty$, that should be equivalent to
M-theory on $T^d$ in the
infinite momentum frame.
However, already for $d=6$ there are problems in implementing this procedure --
in the limit of the D6-brane considered in [\Sen,\Seib], it appears that there
may be difficulties in
decoupling gravity. For $d>6$, there are further problems in even considering
systems of $N$
D-branes: one cannot have arbitrary numbers of D7-branes, the D8-brane requires
the modified type
IIA theory with a mass term [\bergy] and D9-brane backgrounds are inconsistent
except in an
orientifold construction, in which case the number of D9-branes is fixed.

For  low $d$, then,
 the matrix theory for M-theory on $T^d$ is given by a limit of the
world-volume theory of $N$
Dd-branes wrapped around $T^d$.
 For $d \le 3$
this gives
$d+1$ dimensional SYM on $\tilde T \times \R$. For $d=4$, the relation between
the D4-brane and the
M5-brane leads  to the effective world-volume dynamics of the M5-brane, namely
the  5+1 dimensional
(2,0) supersymmetric self-dual tensor theory on $ T^5\times \R$. The type IIA
string coupling is
related to the size of the 11th compactified dimension of   M-theory, while the
4+1
dimensional SYM coupling constant is related to the size  of the extra
dimension for the
6-dimensional tensor theory compactified on a circle. Thus on taking 4+1
dimensional SYM to strong
coupling, a tower of  states become light; these can be interpreted as
Kaluza-Klein modes for a 5+1
dimensional theory  compactified on a circle which becomes large as the
coupling does. In this case,
the states becoming light are the solitons in $4+1$ dimensions arising from
Yang-Mills instantons in
$4$ Euclidean dimensions [\rozali-\MatTor].

For $d=5$, the D5-brane is related by $SL(2,\Z)$ duality to the (solitonic)
 NS 5-brane (or F5-brane) of type IIB, the
world-volume theory for which is the (1,1) supersymmetric non-critical string
proposed in
[\MatTor] whose low energy effective theory  is 5+1 dimensional  SYM. For
$d=5$, the matrix
theory is then (1,1) string theory on $ T^5\times \R$, which is T-dual to the
(2,0)  string theory
on $ T^5\times
\R$. As the M-theory torus decompactifies from $T^5$ to $T^4\times \R$, the
tension of the (2,0)
string becomes infinite so that it reduces to the (2,0) tensor field theory, as
required.

For $d=6$, the D6-brane arises from the Kaluza-Klein monopole of M-theory (i.e.
the solution
$\R^{6,1} \times N$ where $N$ is self-dual Taub-NUT space) [\town] which can be
interpreted as a
6-brane of M-theory [\GravDu]. The corresponding matrix theory is
then the world-volume theory of $N$ G6-branes of M-theory on $T^6\times \R$, in
the large $N$ limit.
The low-energy limit of this is 6+1 dimensional SYM, but the theory must also
contain extended
objects that reduce to the strings of the $d=5$ case in the decompactification
limit. As we shall
see, the theory contains membranes  [\mem] which do precisely this. However,
there is
a difference between this and the $d=5$ case: for $d=5$ we obtain a string
theory, but
for $d=6$ the membranes are not perturbative states, and so the theory is not a
perturbative  membrane theory. Thus the situation is similar to that of
M-theory, which also contains membranes, but is not a perturbative membrane
theory.

For $d\ge 7$, it is hard to generalise this picture due to the problems with
having arbitrary
numbers of Dd-branes.
For $d=8$, the D8-brane has been conjectured to arise from a  9-brane of
M-theory
[\GravDu]. This would suggest that the  $8+1$ dimensional SYM arises from a 9+1
dimensional theory,
and that the matrix  theory is this 9+1 dimensional theory, that gives the
strong-coupling limit of
$N$ D8-branes on
$T^8\times
\R$. Further evidence in favour of this will be presented later.
For $d=7$, the D7-brane   arises
from a Kaluza-Klein monopole of M-theory on a $T^2$ that shrinks to zero size,
where one circle is
transverse and one longitudinal (or from the corresponding F-theory
construction).
 If the M9-brane conjecture is correct, and if a D8-brane arises from
its double dimensional reduction on a circle, then the  D7-brane arises
from an M9-brane on $T^7\times T^2$ in the limit in which the $T^2$ shrinks to
zero size.
Understanding the matrix theory in this case would require a better
understanding of these
formulations of the D7-brane at finite coupling. The case $d=9$ will be briefly
discussed in
section 5.

Toroidally compactified M-theory has a remarkable limit    whose low energy
effective description is
in terms of $d+1$ dimensional SYM.
At least for $d\le 5$, this limit gives a consistent interacting matter theory
which is SYM for $d\le 3$ and gives unexpected new theories for $d=4,5$.
For $d\ge 6$, the limiting theory should still
have an effective SYM description, but there are many unresolved issues, such
as whether there is a
limit in which gravity and the bulk fields decouple.

 The purpose of this paper is to use the expected U-duality
symmetries to try and learn something about the
effective SYM theories that emerge in these limits
of toroidally compactified
M-theory, and then use the effective SYM theories to learn about matrix theory,
assuming it exists
for $d\ge 6$.
It should be emphasized that the considerations in this paper are purely
kinematical and concern the BPS states of the low energy effective
braneworld-volume  theories, and are independent of whether the matrix limits
actually  exist for $d\ge 6$.
The usual analysis is for rectangular tori without background fields; U-duality
requires general backgrounds with arbitrary torus metrics and background
fields.
The full matrix theory is expected to have \lq manifest' U-duality symmetry.
For $d\le 3$, the matrix theory is just the $d+1$ dimensional SYM theory and
the
U-duality is indeed manifest. For $d>3$,
the $d+1$ dimensional SYM theory is only an effective description
and the U-duality is not expected to be manifest in the SYM theory.
Indeed,
U-duality is manifest in the
6-dimensional
self-dual tensor and string theories which are the matrix theories for $d=4,5$,
but it is not a
symmmetry
of the $d+1 $ dimensional SYM theory.
Nevertheless, there is a remnant of U-duality in the SYM theory: it must couple
to all the moduli
that are related to the torus metric by U-duality.
Thus U-duality gives important information about the terms that occur in the
SYM effective action,
and which terms in the D-brane action survive in the \lq matrix theory limit'.
This in turn gives information about the spectrum of BPS branes of the matrix
theory, since these
must arise as solitonic branes of the SYM theory; e.g. it implies the presence
of
$d-4$ branes, giving the  0-branes in 4+1 dimensions that correspond to
Kaluza-Klein modes of the
matrix theory, and the
solitonic strings in 5+1 dimensions which correspond to the fundamental strings
of the matrix
theory.

 For example,
M-theory on $T^3$ should have $SL(3,\Z)\times SL(2,\Z)$ U-duality, implying
that the
corresponding $3+1$ dimensional  SYM theory should have $SL(2,\Z)$  symmetry.
The appropriate limit of the D3-brane action includes a $\theta$-angle
coupling, and we
learn that this is essential for the
 $SL(2,\Z)$ S-duality, and that the theta-angle and
the coupling constant    take values in an $SL(2)/U(1) $ coset space.
Of course, all this was already well-known for 4-dimensional SYM,
but a similar analysis gives interesting predictions for higher values of $d$.
In particular, we will learn that for $d=6$, there is a generalised theta-angle
(for certain \lq instantonic 3-branes') that combines with the coupling
constant
to form a complex field taking values in an $SL(2)/U(1) $ coset space, and
which is acted on by
an $SL(2,\Z)$ duality symmetry.
In this way, useful information will be learned about the effective SYM
description of these
limiting theories, whenever the limits exist.

In each case, the SYM action that is found depends on the metric  and
certain constant anti-symmetric tensors on the  torus.
These can be thought of as generalised coupling constants and arise from the
expectation values of
background supergravity fields [\Ym].
For example, for $d\ge 3$, the $D=d+1$ dimensional action includes the coupling
to a $D-4$ form,
whose Hodge dual is a 4-form $X$, and is of the form
$$ {1\over 4 g^2} \tr F_{mn}F^{mn} +{1\over 2}X^{mnpq}\tr F_{mn} F_{pq}
\eqn\ymt$$
Some of the consequences of such terms were considered in [\Ym].

\chapter{U-Duality}

M-theory compactified on $T^d$  has U-duality group $E_d(\Z)$ and scalars
taking values in
$E_d/H_d$, where $H_d$ is the maximal compact subgroup of $E_d$.  In each case,
SYM on
$\tilde T^d\times \R$ has a manifest $SL(d,\Z)$ symmetry acting naturally on
the torus,
and the torus metric moduli lie in $\R \times SL(d)/SO(d)$.
The SYM action includes the kinetic term, depending on the torus metric, and
couplings to all
the background fields related to this by $E_d$ duality, so that it couples to
all moduli in $E_d/H_d$.
As the $SL(d)$ subgroup of $E_d$ acts through torus diffeomorphisms,
decomposing
$E_d/H_d$ into $SL(d)$ representations  will tell us the tensor structure of
the
SYM coupling constants.
 We will
consider each value of $d$ in turn. For $d\le 5$, the results of [\Mat-\MatTor]
are recovered, but
the analysis makes interesting predictions for $d>5$.

\subsection {$d\le 3, E_3=SL(3)\times SL(2)$}

The moduli space $SL(3)/SO(3)\times SL(2)/U(1)$ corresponds to metrics on
$T^3$, plus one
additional scalar, the $\theta$-angle $C_0$, which combines with the torus
volume (which determines
the SYM coupling constant $g$) to take values in $SL(2)/U(1)$.
The lagrangian includes the terms
$$L \sim {1\over g^2} \tr F\lll *F + C_0 \tr F\wedge F
\eqn\thr$$
and the geometric
$SL(3,\Z)$ acting on the torus combines with the $SL(2,\Z)$ S-duality, so that
the full $E_3(\Z)$
U-duality is manifest.  The U-duality is also
manifest for $d<3$, as follows immediately  by taking suitable limits.

\subsection {$d=4, E_4=SL(5)$}

The parameter space of the theory is $SL(5)/SO(5)$, while the moduli of the
4-torus are in
$\R\times SL(4)/SO(4)$; the  extra  parameters in $SL(5)/SO(5)$ are in a {\bf
4} of $SL(4)$ and
so are associated with a 1-form $C_1$ on $\tilde T^4$.
(A 3-form coupling through $C_3\lll F$ would have given the same counting, but
it will be argued
in section 3 that it is the 1-form that gives the correct SYM theory, as used
in
[\rozali-\MatTor].) The lagrangian for
$4+1$ SYM then includes the terms
$$L \sim {1\over g^2} \tr F\lll *F + {C_1} \wedge \tr F\wedge F
\eqn\fou$$ However, {\it any} $SL(4)\times \R$ subgroup of $SL(5) $ can be
associated with a 4-torus
in this way, so that different 4-tori emerge from different limits of the
moduli space. This suggests
that the full theory be formulated on a 5-torus, and be given by a 5+1
dimensional field theory on
$\tilde T^5
\times \R$. It was argued in [\rozali-\MatTor] that this is indeed the case,
and that the matrix
theory is given by the (2,0) supersymmetric self-dual tensor theory on $\tilde
T ^5\times \R$. This
theory has an ultra-violet fixed point [\MatTor], and the scale invariance then
implies that the
theory is independent of the volume of the $\tilde T^5$, explaining the absence
of an extra $\R$
factor in the moduli space
$SL(5)/SO(5)$.
The self-duality of the tensor theory implies that it has no adjustable
coupling constant. The
expected  4+1 SYM emerges in the appropriate limits [\rozali-\MatTor].

\subsection {$d=5, E_5=SO(5,5)$}

The parameter space of the theory is $SO(5,5)/SO(5)\times SO(5)$, while the
moduli of the 5-torus
are in
$\R\times SL(5)/SO(5)$; the  extra  parameters in $SO(5,5)/SO(5)\times SO(5)$
are in a {\bf 10} of
$SL(5)$ and so are associated with a 2-form $C_2$ on $\tilde T^5$. (A 3-form on
$\tilde T^5$ would
also
be in the  {\bf 10}, but the D5-brane does not couple to a 5-form.)
The lagrangian for
$5+1$ SYM then includes the terms
$$L \sim {1\over g^2} \tr F\lll *F  + {C_2} \wedge \tr F\wedge F
\eqn\fiv$$
This is part of the low-energy effective action of the (1,1) supersymmetric
string theory in 5+1
dimensions proposed in [\MatTor]. The
  strings coupling to $C_2$ arise as solitons of the low-energy effective
theory, which in this case
correspond to   YM instantons.  It was proposed in [\brs,\MatTor] that M-theory
on $T^5$ corresponds
to this string theory on $\tilde T^5\times \R$.
The matrix theory limit requires going to strong string coupling, so that it is
useful
to first perform an $SL(2,\Z)$ duality transformation which takes the D5-brane
to the NS 5-brane
so that the limit is now one of weak string coupling. The theory emerging in
this limit\
is a 6-dimensional (1,1) supersymmetric  non-critical string theory with
the SYM theory as its zero slope limit. For the D5-brane, $C_2$ is the RR
2-form while for the NS
5-brane it is the NS 2-form.
 This theory is related by T-duality to the (2,0) string on a
5-torus [\MatTor] whose low-energy limit is the tensor theory that arose for
$d=4$, so that either
string theory on a 5-torus can be used to describe M-theory on $T^5$.

\subsection {$d=6, E_6$}

This is the first  case in which we learn something new. The parameter space of
the theory is $E_6/USp(8)$, while the moduli of
the 6-torus are in
$\R\times SL(6)/SO(6)$.
$E_6$ has a maximal subgroup $SL(6)\times SL(2)$,  under which ${\bf 78}
\rightarrow {\bf (1,3)+(35,1)
+(20,2)}$. The  parameter space $E_6/USp(8)$ then decomposes into $SL(6)/SO(6)
\times SL(2)/U(1)$,
plus an extra 3-form $C_3$ on $\ti T^6$ in the {\bf 20} of $SL(6)$.
The 3-form couples through a $ {C_3}   F^2$ term, but the coupling to the extra
scalar
$\chi$ that combines with the coupling constant $g$ (or, equivalently, the
torus volume)
to parameterise the coset space $SL(2)/U(1)$ is more problematic. It will be
argued  in section 3
that this coupling is through a term $(DX)^3 F^2$, where the $X$ are the three
adjoint scalars in the
$6+1$ dimensional SYM multiplet, and $DX=dX+[A,X]$.
 The resulting lagrangian for $6+1$ SYM then includes the terms
$$L \sim {1\over g^2} \tr  F\lll *F + {C_3} \wedge \tr ( F\wedge F )
+\chi \Str \,( DX \wedge DX \wedge
DX)
\wedge F\wedge F
\eqn\six$$
where Str denotes the symmetrised trace.
The term   $ \chi (DX)^3 F^2$ involves an \lq axionic' scalar $\chi$ and is
obtained from 10
dimensions by the reduction of the term
$\chi \tr F^5$, so that $\chi$ is a generalised $\th$ angle.
Such a $\chi \tr F^5$ term arises formally as part of the D9-brane action.

 In this theory, YM instantons
give solitons of the low-energy theory that correspond to membranes coupling to
$C_3$. Thus the 6+1
dimensional theory contains membranes and gives rise  to the  strings of the
(1,1) theory in 5+1
dimensions on double dimensional reduction,
and to membranes which are expected to be Dirichlet branes of the non-critical
string theory
on simple dimensional reduction.
The \lq $\theta$-angle' $\chi$ couples to certain instantons which will be
discussed in section 4.

\subsection {$d=7, E_7$}

 The parameter space of the theory is $E_7/SU(8)$, while the
moduli of the expected 7-torus are in
$\R\times SL(7)/SO(7)$. The 70 parameters in $E_7/SU(8)$
  then decompose  into the {\bf 27+1} moduli of $T^7$ in $\R\times
SL(7)/SO(7)$, together with a
{\bf 35 }and a {\bf 7} of SL(7). The {\bf 35 } could be
a 4-form or a 3-form on
$T^7$, while
  the {\bf 7} could be a vector or a 6-form. The D7-brane does not couple to
a vector or 3-form, but
with a 4-form and a 6-form, one can write the following
  action for $7+1$ SYM
$$S= \int {1\over g^2} \tr F \lll *F  + {C_4} \wedge \tr F\wedge F +C_6 \lll
\tr  F
\eqn\sev$$
Other possibilities for the action will be discussed in section 3. For the
action \sev, the YM
instantons   correspond to 3-branes coupling to $C_4$.

In fact, $E_7$  contains $SL(8)$ as a
maximal subgroup and under this the parameter space decomposes into
$SL(8)/SO(8)$ together with a $35^+$ of $SL(8)$, corresponding to a self-dual
4-form. The situation
is then similar to that for $d=4$. Any $SL(7)$ subgroup of $SL(8)$ is
associated with a 7-torus,
suggesting that the theory could be an 8+1 dimensional theory on $T^8\times
\R$. Moreover, this
theory should be  scale invariant, to account for the absence of a modulus for
the volume of the
8-torus. The 8+1 dimensional  theory should  contain 3-branes coupling to the
self-dual 4-form on
$T^8$. However, it is problematic to obtain a self-dual 4-form on $T^8$ from a
covariant field on
$8+1$ dimensions.
Moreover, for a standard  Kaluza-Klein picture to apply, one would expect the
{\bf 7} to
correspond to a vector, not a 6-form, with the  vector interpreted as a
gravi-photon coupling to
0-branes that become light in the decompactification limit.

\subsection {$d=8, E_8$}

Finally, in this case the parameter space   is $E_8/SO(16)$, while the moduli
of the expected
8-torus are in
$\R\times SL(8)/SO(8)$. The 128 moduli in $E_8/SO(16)$ correspond to the {\bf
35+1} moduli of $\ti
T^8$,  plus parameters in the {\bf 56,28,8} representations of
$SL(8)$.
The {\bf 56} could
correspond
to a 5-form  or 3-form on $\ti T^8$,
with the coupling
$C_5\lll F^2$ or $C_3\lll F^3$, while
the {\bf 28} could
correspond
to a 2-form  or 6-form,
with a coupling $C_2\lll  DX \lll F^3$ or $C_6 \lll DX \lll F$.
The {\bf 8} could
correspond
to a 1-form or 7-form on $\ti T^8$,
with the coupling
$A_1 \lll *J$ or $C_7 \lll F$,
where
$J$ is the conserved current $J_m= F_{mn}D^nX$.
If a 5-form, 2-form and 7-form are chosen, for example, an action of the form
$$S= \int \Str \left[
{1\over g^2}  F \lll *F  + {C_5} \wedge  F\wedge F +C_7 \lll   F + C_2 DX F^3
\right]
\eqn\abc$$
can be written down;
  another possibility with background $1,5$ and $6$ forms is
$$
S= \int \Str \left[
{1\over g^2} F\lll *F + C_5\lll F\lll F +\ww _6 \lll  DX\lll  F + A_1 \lll DX
\lll
*F
\right]
\eqn\eig$$
and other possibilites will be considered in section 3.

However, $E_8$ in
fact contains $SL(9)$ as a maximal subgroup.
As before, any $SL(8)$ subgroup of $SL(9)$ can be associated with an 8-torus,
suggesting that
  the theory could be a scale-invariant 9+1
dimensional theory on $T^9\times \R$, in which case the moduli correspond to
the {\bf 44} constant
volume metrics on $T^9$ in $SL(9)/SO(9)$, together with a 6-form $C_6$ in the
{\bf 84} of $SL(9)$. A
  9+1 dimensional SYM lagrangian can be written with a coupling to a metric and
6-form,
containing the following terms
$$L \sim {1\over g^2}\tr ( F\wedge  *F )+ {C_6} \wedge \tr (F\wedge F)
\eqn\abc$$
The supersymmetric lagrangian is then that of $D=10$ SYM, plus a topological
term $C_6\lll F^2$.
Remarkably,
the reduction of this to 8+1 dimensions gives the candidate lagrangian \eig\
(at least to linearised
order).
Indeed the charge $J_0$ with respect to the background Maxwell field $A_1$ is
just the momentum in
the compactified direction. Thus   if the action is \eig, there is the
possibility that 0-branes of
the 8+1 dimensional theory carrying the
charge $J_0$ could be reinterpreted as Kaluza-Klein modes for a  compactified
10-dimensional theory.

\subsection {Summary}

For each $d$, the low-energy effective action is SYM on $\ti T^d\times \R$. The
coupling constants
consist of (i) the torus metric and SYM coupling constant (which can be
absorbed into the volume of
the torus), taking values in $\R\times SL(d)/SO(d)$ (ii) a $d-3$ form for $d
\ge 3$
(or its dual, a 4-form)
(iii) a $d-6$
form for $d \ge 6$ (or its dual, a 7-form) and (iv) a
1-form for $d = 8$. Thus for $d<8$ we obtain  $d+1$ dimensional SYM coupling to
a
metric, a 4-form and a 7-form.
The $d-3$-form couples through a term of the form
$$ C_{D-4}\lll F \lll F
\eqn\ymta$$
which can be written in terms of the 4-form $X=*C_{D-4}$ as in \ymt.
Instantons in 4 Euclidean dimensions give rise to $d-4$ brane solitons of the
SYM theory
that carry charge with respect to the   potential $C_{D-4}$.
For $D=6$, the 1-brane solitons of the SYM theory are identified with the
fundamental strings of
the 6-dimensional string theory
 and for $D=5$, the 0-branes are Kaluza-Klein modes of the 6-dimensional tensor
theory.
For $d=7$, there is either an extra vector which would couple to 0-branes, or
an extra 6-form
which would couple to 5-branes.
For $d=8$ there is a vector or 7-form, coupling to 0-branes or 6-branes, and
a 2-form or 6-form coupling to strings or 5-branes.
 Furthermore, U-duality suggests that for $d=4,7,8$ the full theory
could in fact be a scale-invariant theory in
$d+2$ dimensions, and these are precisely the cases in which there may be
0-brane solitons which
could represent Kaluza-Klein modes.
It is clearly important to find further restrictions on the cases where the
group theory does not
give a unique prediction, and this will be done in the next section.

\chapter{D-Brane Actions and U-Duality}

The matrix model for M-theory on $T^d$ or a Dd-brane wrapped on $T^d$
is a SYM theory coupling to the metric on $T^d$, so that the metric provides
coupling constants of
the theory, taking values in $GL(d)/SO(d)$. If the U-duality group $E_d$ is to
be realised
on the theory, then the SYM should couple to all the scalars related to
the metric coupling constants by U-duality, so that the SYM should couple to
scalars in
$E_d/H_d$ where $H_d$ is the maximal compact subgroup of $E_d$. Moreover, the
conjecture  of [\Sus]
suggests that
in fact the SYM theory
at finite $N$ realises the U-duality group $E_{d+1}$.
The D-brane couples to all scalars in
$E_{d+1}/H_{d+1}$, and the
 matrix theory of [\Sus] is obtained as a limit of this D-brane theory
[\Sen,\Seib].
 We will now investigate these couplings.

 The D-brane action for a single $p$-brane in a supergravity background with
constant fields is a
$D=p+1$ dimensional action including the terms [\Dbrane,\Doug,\polch]
$$\eqalign{
S=  &\int d^D x \, e^{-\Phi} \sqrt{ \det {(g_{mn}+{\cal F}_{mn})}}+
\int  \left[  C_D +C_{D-2}{}\lll {\cal F}
+C_{D-4}{}\lll {\cal F} \lll {\cal F} \right.
\cr &+
\left. C_{D-6}{}\lll {\cal F} \lll {\cal F}\lll {\cal F}
+...
+C_{D-2r}{}\lll {\cal F}^r \right]
\cr}
\eqn\Dact$$
where ${\cal F} _{mn} =F_{mn} -b_{mn} $,
$r$ is the integer part of $D/2$,
$g_{mn}, b_{mn}$ are the pull-backs of the metric $G_{MN}$ and   the NS-NS
2-form gauge field
$B_{MN}$
$$g_{mn}=G_{MN} \pa _{m } X^{M }\pa _{n} X^{N}  , \qquad b_{mn}=B_{MN} \pa _{m
} X^{M }
\pa _{n} X^{N}
\eqn\gis$$
 and the $C_r $ are $r$-forms arising from the background
expectation values of the pull-backs of RR gauge fields,
so that
$$C_r=C_{M_1.... M_r} \pa _{m_1} X^{M_1}
...
\pa _{m_r} X^{M_r} d\ss ^{m_1} \lll ...\lll d \ss ^{m_r}
\eqn\cis$$
where $M,N=0,1,...9$ are space-time indices and $m,n=0,1,...,p$ are $p$-brane
world-volume indices.
For $r>4$, the $C_r$ are the dual potentials; the field equation for
$C_r$ with $r \le 4$ is of the form $dG=0$, where $G= *dC_r+...$ is the $9-r$
form field strength,
and the dual potential $  C_{ 8-r} $ is the potential for $G$, $G=dC_{8-r}$
[\Dbrane].
Similar actions    arise in matrix theory.
{}From the  point of view of the SYM theory, the  forms $C_r$ are  coupling
constants.
On going to static gauge, the $X^M$ are split into coordinates $X^m$, which are
identified
with the $\ss^m$ so that $\pa _m X^n= \dd_m {}^n$, together with the transverse
coordinates
$X^\aa $ where $\aa = 1,..., 9-p$.
Then  the tensor $C_{M_1.... M_r}$ splits into a set of forms
$C_{m_1...m_t  \aa _1 ...\aa_{r-t} }$ for  various $t$, and $C_r$ \cis\ becomes
$$\eqalign{C_r=&\sum _t
C_{m_1...m_t   \aa _1 ...\aa_{r-t} } \pa _{m_{t+1}} X^{\aa_1}
...
\pa _{m_r} X^{\aa_{r-t}} d\ss ^{m_1}\lll  ... \lll d \ss ^{m_r}
\cr
=&
\sum _t
(C_t)_{   \aa _1 ...\aa_{r-t} }d X^{\aa_1}
\lll dX^{\aa_2}\lll
...\lll
d X^{\aa_{r-t}}
\cr}
\eqn\abc$$
where the $t$-form $C_t$ is defined by
$$
(C_t)_{   \aa _1 ...\aa_{r-t} }=
C_{m_1...m_t   \aa _1 ...\aa_{r-t} }  d\ss ^{m_1}\lll ... \lll d \ss ^{m_t}
\eqn\abc$$
Thus   \Dact\
contains terms such as
$$
(C_t)_{   \aa _1 ...\aa_{s} }d X^{\aa_1}
\lll dX^{\aa_2}\lll
...\lll
d X^{\aa_{s}}\lll {\cal F}^m
\eqn\cter$$
with $D=t+s+2m$.
Similarly, $g_{mn},b_{mn}$ become
$${\eqalign{
g_{mn}&=G_{mn}
+2G_{\aa (m}\pa _{n)}X^\aa
+
G_{\aa \nn} \pa _{m } X^{\aa }\pa _{n} X^{\nn} ,
\cr
 b_{mn}&
=B_{mn}
+2B_{\aa (m}\pa _{n)}X^\aa
+
B_{\aa \nn} \pa _{m } X^{\aa }\pa _{n} X^{\nn}
\cr}}
\eqn\gist$$

For $N$ D-branes, the gauge symmetry becomes $U(N)$ and the gauge fields and
the
transverse scalars $X^\aa$ take values in the adjoint representation of $U(N)$.
The derivative $dX^\aa = \pa _{m} X^{\aa}
d\ss ^m$ is replaced by the covariant derivative
$DX^\aa = D _{m} X^{\aa}
d\ss ^m$, where
$$ D_m X^\aa = \pa _m X^\aa + [A_m, X^\aa]
\eqn\abc$$
Terms in \Dact\ that are independent of $X$
become of the form $C_s \tr {\cal F}^t$
where now
$$ {\cal F} _{mn} =F_{mn} -b_{mn} \unit
\eqn\abc$$
but \cter\ has a number of possible
  non-abelian generalisations. The simplest
guess would be
$$
(C_t)_{   \aa _1 ...\aa_{s} }{\rm  Str} \left[
D X^{\aa_1}
\lll DX^{\aa_2}\lll
...\lll
d X^{\aa_{s}}\lll {\cal F}^m
\right]
\eqn\abc$$
 and this is supported by [\Cal].

The D-brane action \Dact\ defines the coupling of the D-brane to all of the 128
bosonic degrees of
freedom of the type II string, consisting of $G_{MN},B_{MN},\fff$ together with
the RR gauge fields;
these can be thought of as the generalised coupling constants of the SYM
theory. For
compactification of the type II theory on
$T^d$ to
$\R^{9-d,1}$, the 128 bosonic degrees of freedom can be split into fields on
$\R^{9-d,1}$ of various spins.
 In particular, the scalars are the moduli for type
II on $T^d$, corresponding to the metric moduli plus the moduli of flat
anti-symmetric tensor
gauge fields on $T^d$. These parameterise the
coset space $E_{d+1}/H_{d+1}$ where $H_{d+1}$ is the maximal compact subgroup
of $E_{d+1}$.

 In [\Sus] it was conjectured that M-theory on a
null circle of finite radius $R_{11}$ in the discrete light-cone gauge is
described by a matrix model
with $U(N)$ gauge symmetry for finite
$N$. In [\Sen,\Seib], this was related to a limit of  the theory
of $N$ D-branes, so that
  M-theory on $S^1\times T^d$ (with null $S^1$) should be described by $U(N)$
gauge theory
(for finite $N$) on
$\ti T^d
\times \R$.
The Dd-brane
theory has $E_{d+1}(\Z) $ symmetry instead of the $E_{d}(\Z) $ symmetry
expected from the old matrix model conjecture and
 it should couple to fields in the coset space $E_{d+1}/H_{d+1}$ (at least
before taking the limits
of [\Sen,\Seib]), not just the coset $E_{d}/H_{d}$. The coupling to these
fields is precisely what is
found in the D-brane action. The matrix model action for infinite $R_{11}$ is
obtained by a
truncation to the terms coupling to  a  $E_{d}/H_{d}$ subspace of
$E_{d+1}/H_{d+1}$. In particular, this
truncation involves dropping the dilaton coupling, as the matrix model limit
involves the weak
string coupling limit
$\ti g \rightarrow 0$. In the following, we will investigate
 which fields should be dropped in this limit to
obtain a SYM theory in $d+1$ dimensions coupling to all moduli in $E_d/H_d$.
Although the truncations can be found by careful consideration of the matrix
theory limit, it is of interest to see how much can be derived using kinematics
and group theory, as these methods can be applied more generally, including
cases in which the matrix theory limit is not yet understood (and may not
exist, at least in the usual form).

Our conventions are that $X^M$ are the space-time coordinates in ${\cal D}=10$
or ${\cal
D}=11$ dimensions (depending on context) and that on compactification these are
split into
$X^M=(X^\mm ,X^i)$, where
$X^i$ ($i=1,...,d$) are coordinates on $T^d$, and $X^\mm$ are space-time
coordinates
($\mm=0,1,2,..., {\cal D}-d$).
On a $d$-brane, the coordinates can instead be split into
$X^M=(X^m,X^\aa)$ where $X^m$ are longitudinal coordinates ($m=0,1,...,d$)
and $X^\aa$ are transverse ($\aa=1,...,{\cal D}-d-1$), so that for a Dd-brane
wrapped on $T^d$,
$X^m=(X^0,X^i)$ and $X^\mm=(X^0,X^\aa)$.
 Note that
each field that emerges from an $r$-form gauge field $C_r$ on toroidal
reduction can also be regarded
as coming from its dual, $\ti C_{8-r}$. For example, for the type IIB theory
reduced on $T^d$, there
are
$d(d-1)/2$ scalars
$C_{ij}$ that arise from a two-form $C_2$ on $T^d$. In $10-d$ dimensions,  an
$8-d$ form
is dual to a scalar, and the $d(d-1)/2$ scalar degrees of freedom could instead
be thought of
as arising from the dual potential $C_6$, giving the $d(d-1)/2$ {} $8-d$ forms
$C_{\mm_1 ...\mm
_{8-d} i_1...i_{d-2}}$. It is important not to double-count such dual
realisations of the same
degrees of freedom in the following.
When there is an ambiguity, we will choose the dual forms that have local
couplings to the Yang-Mills theory when it is weakly coupled.

\subsection {$d=3$}

Consider first the case $d=3$. The D3-brane wrapped on $T^3$ has an action
which couples to the
constant fields
$G_{ij}, B_{ij}, C_{ij}, C_0, \Phi$ on the torus, transforming in the {\bf
5+1+3+3+1+1} of $SL(3)$ giving 14 scalars, which
parameterise the coset space $E_4/H_4=SL(5)/SO(5)$.
We wish to find the restriction to  scalars in the {\bf 5+1+1} of $SL(3)$
parameterising
$SL(3)/SO(3)\times SL(2)/U(1)$ to obtain the matrix theory for M-theory on
$T^3$. The {\bf 5} clearly
corresponds to the metric, and the weak coupling limit requires that the
dilaton be dropped, so that
the {\bf 1+1} must correspond to the size of the torus and the axion $C_0$. The
SYM coupling
constant is proportional to the volume of the 3-torus, and we learn that $g$
and $C_0$ together
parameterise the coset space
$SL(2)/U(1)$ and that there is an $SL(2,\Z)$ duality symmetry which acts on
them.
The action includes the terms \thr. Thus the matrix theory conjecture together
with the conjectured
U-duality of the 8-dimensional string theory [\HT] predicts the S-duality of
4-dimensional SYM.
Although this S-duality is not a surprise here, this will have interesting
generalisations for higher $d$.

\subsection {$d=4$}

For $d=4$, the D4-brane couples to $G_{ij}, B_{ij}, C_{ijk}, C_i, \Phi$
in the {\bf 9+1+6+4+4+1} of $SL(4)$ giving 25 scalars, which
parameterise the coset space $E_5/H_5=SO(5,5)/SO(5)\times SO(5)$.
To obtain the restriction to $SL(5)/SO(5)$,  we need to truncate to
14 scalars in the  {\bf 9+1+4}
representation of $SL(4)$. This should not include the dilaton, so that the
only choice is between
$C_i $ and $ C_{ijk}$ to give the {\bf 4}. The D4-brane couples to both of
these, through
$C_1 F^2$ and $C_3F$.
For the $d=3$
case, we saw that it was necessary to keep
$C_0$ and throw out
$C_{ij}$, so to recover this case in the appropriate decompactification limit
requires keeping $C_i $ and not $ C_{ijk}$
in this case. We are thus led to the the action given by \fou, with the
coupling constant determined
by the torus volume.

\subsection {$d=5$}

For $d=5$, the D5-brane   couples to
$G_{ij}, B_{ij}, C_{ij},C_{ijkl}, C_0, \Phi$ in the {\bf 14+1+10+10+5+1+1} of
$SL(5)$ giving 42
scalars, which parameterise the coset space $E_6/H_6=E_6/USp(8)$. To obtain the
restriction to
$SO(5,5)/SO(5)\times SO(5)$, we need to truncate to 25 scalars in the {\bf
14+1+10} representation
of $SL(5)$. This should not include the dilaton,
and the volume of the metric should be chosen instead of $C_0$ to give the
extra
singlet since the torus volume was included for $d<5$.
The {\bf 10} could come from either the NS two-form $B_{ij}$  or the RR
two-form $C_{ij}$.
The natural choice is to
  exclude the coupling to $B_{ij}$
because the $d=3,4$ cases did not include it, and these cases should be
recovered in the limit in
which
$T^5$ decompactifies to $T^3\times \R^2 $ or
 $T^4\times \R  $.
Then restricting to $G_{ij},C_{ij}$ gives
the coupling to 25 scalars in the {\bf 14+1+10} representation of
$SL(5)$ parameterising
$SO(5,5)/SO(5)\times SO(5)$,  with  the action given by \fiv. This is the
action for the D5-brane; a
type IIB $SL(2,\Z)$ transformation takes this to the NS 5-brane and replaces
$C_{ij}$ with $B_{ij}$,
to give a coupling
$B\lll F \lll F$, as in [\brs,\MatTor], and it is in this form of the theory
that the
appropriate  matrix theory limit is that of weak string coupling
[\brs,\MatTor] .

\subsection {$d=6$}

For $d=6$, the 6-brane couples to $G_{ij}, B_{ij}, C_{ijk}, C_i, \Phi$
in the {\bf 20+1+15+20+6+1} of $SL(6)$ giving 63 scalars, instead of the 70
that would be needed
for the coset space $E_7/SU(8)$.
The remaining scalars arise from the dualisation of anti-symmetric tensor gauge
fields.
The bosonic spectrum of type IIA compactified on $T^6$ contains,
 in addition to the 63 scalars,
{\bf 1+6} 2-forms from $B_{\mm \nn} $ and $C_{\mm \nn i}$ and one 3-form
$C_{\mm \nn \rr}$, together with a metric $G_{\mm \nn}$ and {\bf 6+6+15+1} =28
vectors
from $G_{\mm j}, B_{\mm j}, C_{\mm jk}, C_\mm $.
For IIA on $T^6$, the dimensionally reduced theory is 4-dimensional and
in 4 dimensions 2-forms are dual to scalars, so that the 63 scalars together
with the extra 7
scalars arising from dualising the 2-forms parameterise the 70-dimensional
coset space $E_7/SU(8)$.
Equivalently, these extra {\bf 1+6} scalars can be viewed as coming from the
coupling to the
   components $C_{ijklm}$ of the potential $C_5$ (dual to the 3-form potential
$C_3$) and to
$B_{ijklmn}$ (dual to the NS 2-form).  Any of the scalars arising from an
$r$-form gauge field can
equivalently arise as a 2-form obtained from  the dual $8-r$ form potential.
For example, the {\bf 6} scalars $C_i$ could be represented by the {\bf 6}
dual
2-forms $C_{\mm \nn ijklm}$ obtained from $C_7$, the dual to $C_1$.
There are similar dual forms of the compactifications for other $d$.

Restricting to $G_{ij}, C_{ijk}$ gives the coupling to {\bf 20+1+20} =41
scalars.
For M-theory on $T^6$, the compactified theory is 5-dimensional
with 42 scalars in $E_6/USP(8)$  in the {\bf 20+1+20+1} of $SL(6)$, so that an
extra singlet is
needed.
There are no other singlet scalars to choose from, and
the 2-forms do not give scalars in 5-dimensions.
However,   in 5 dimensions 3-form gauge
fields are dual to scalars.
 Dualising the single 3-form $C_{\mm \nn \rr}$
gives an extra scalar, to give the 42 scalars parameterising $E_6/USp(8)$.
This fits in with the M-theory picture, where on compactification of M-theory
on $T^6$, the
{\bf 20+1+20+1} scalars come precisely from $G_{ij}, A_{ijk}, A_{\mm \nn \rr}$
where here $A_{MNP}$ is the 3-form of 11-dimensional supergravity.
Thus both M-theory and matrix theory give 41 scalars
(taking values in $G/H$ where
$G=GL(6,\R)/SO(6)\times \R^{20}$
[\PL]) and a 3-form.
(In fact, the D6-brane also couples to 3-forms $C_{\mm \nn \rr ij},C_{\mm \nn
\rr ijkl}$ arising from
$C_5,C_7$, but these are both in the {\bf 15} of $SL(6)$ and so do not give
singlets.)
 Thus we learn that  the matrix model for M-theory on $T^6$ in the limit
$R_{11} \to \infty$
has as its low-energy effective theory SYM coupled to the scalars $G_{ij},
C_{ijk}$ and the 3-form
$C_{\mm \nn \rr}$.
The coupling to $C_{\mm \nn \rr}$ follows from the D-brane action \Dact\ (with
abelian gauge group),
and is of the form
$$C_{MNR}  (dX^M \lll dX^N \lll dX^R \lll F\lll F)
\eqn\abc$$
The 10 $X^M$ split into  three transverse $X$'s, $X^\aa$ ($\aa, \bb =1,2,3$)
and the 6+1 longitudinal $X$'s,  $X^m$, which
can be identified with the world-volume coordinates on
going to static gauge, so that the action becomes
$$C_{\aa \bb \gg}  (dX^\aa \lll dX^\bb \lll dX^\gg \lll F\lll F)
\eqn\abc$$
Now $C_{\aa \bb \gg}=\chi \ee _{\aa \bb \gg}$ for some $\chi$, and the SYM
coupling becomes
$$\chi \ee _{\aa \bb \gg}  (dX^\aa \lll dX^\bb \lll dX^\gg \lll F\lll F)
\eqn\abc$$
The natural candidate for  the
  non-abelian generalisation of this is [\Cal]
$$\chi \ee _{\mm \nn \rr}\Str (DX^\mm DX^\nn DX^\rr F^2)
\eqn\cert$$
This gives the action \six.

\subsection {$d=7$}

For $d=7$, the D7-brane   couples to
$G_{ij}, B_{ij}, C_{ij},C_{ijkl}, C_0, \Phi$ in the {\bf 27+1+21+21+35+1+1} of
$SL(7)$ giving 107
scalars.
In addition there are {\bf 7+7+7 }=21 vectors $G_{i\mm}, B_{i\mm}, C_{i\mm}$.
In the 3-dimensional theory obtained by
compactifying the type IIB theory on $T^7$, vectors are dual to scalars and
dualising these 21 vectors  give 21 additional scalars and a total of
 $107+21=128$ scalars parameterising $E_8/SO(16)$. The {\bf 7+7 } vectors $
B_{i\mm}, C_{i\mm}$
could instead be given as scalars arising from the components
$B_{ijklmn},C_{ijklmn}$
of the dual potentials $B_6,C_6$.
M-theory on $T^7$ should couple to 70 scalars in the coset space $E_7/SU(8)$
and as we saw in section
2, the group theory implies that these should  be in the {\bf 27+1+35+7} of
$SL(7)$.
The scalars in the {\bf 27+1+35 } representation should come from $G_{ij},
C_{ijkl}$
but there is more than one possibility for the {\bf 7}.
The simplest is that these should arise from the scalars $C_{ijklmn}$, through
the coupling $C_6\lll F$. However,
in 4 dimensions, 2-forms are dual to scalars and the
 missing degrees of freedom could arise from {\bf 7} 2-forms.
The potentials $B_2,C_2,C_4, C_6,B_6,C_8$ give rise to {\bf 1+1+21+35+35+7}
2-forms
(e.g $C_4$ gives the
 21 2-forms
$C_{ij\mm \nn}$).
These do not give rise to propagating degrees of freedom in 3 dimensions but
are dual to scalars in 4
dimensions. Thus the only possibility is through the coupling to the components
$C_{ijklmn\mm \nn}$
of
$C_8$,  which couples to the 2 transverse scalars $X^\aa$ through
the coupling
$\ww _6 \lll DX \lll DX$ where $C_{ijklmn\aa \bb}=\ww_{ijklmn}\ee_{\aa \bb}$.
The fact that M-theory on $T^7$ gives 63 scalars
from constant metric $G_{ij}$ and  3-form gauge fields $A_{ijk}$ on   $T^7$
and  7 space-time 2-form gauge fields $A_{i \mm \nn}$ might be construed as
evidence in favour
of the coupling to $C_8$, which gives the coupling to 7 2-forms.
In both cases, there is a 6-form which couples to a 5-brane inside the 7-brane,
as will be
discussed in section 4.

\subsection {$d=8$}

For $d=8$, the D8-brane couples to the background fields $G_{ij}, B_{ij}$,
$C_{ijk}, C_i, \Phi$,
which transform
in the {\bf 35+1+28+56+8+1} representations of $SL(8)$,
giving 129 moduli. Alternatively, the {\bf  56+8 }  could be thought of as
arising form the dual
potentials
 $C_{ijklm},C_{ijklmnp} $.
 In three dimensions, vectors are dual to scalars and there are vectors
in the {\bf 8+8+1+28+70+28+1+56} of $SL(8)$
from
 $G_{\mm j}, B_{\mm j},C_\mm, C_{\mm jk},C_{\mm jklm},C_{\mm jklmnp} ,C_{\mm
ijklmnpq},
B_{\mm ijklm} $.
M-theory on $T^8$ gives a 3-dimensional theory with scalars in the
  coset $E_8/SO(16)$, which has 128 scalars which are in the
{\bf 35+1+28+56+8 }  of $SL(8)$.
The {\bf 35+1  } should arise from $G_{ij}$, while the {\bf 56} could arise
from $C_{ijk}$,
$C_{ijklm}$ or
$B_{\mm ijklm} $. The fact that all cases with $d<8$ included the coupling
$C_{d-3}\lll F\lll F$
suggests that the coupling to $C_{ijklm}$ is required here.
The {\bf 28 } could arise from $B_{ij}$ or $C_{\mm jk}$ or $C_{\mm jklmnp}$;
however, the
 coupling to $B_{ij}$  can be excluded because it didn't occur for $d<8$. Here
there is one
transverse scalar $X$ so in static gauge $\aa$ takes only one value, so writing
$\ww_{  jk }=C_{\aa
jk }$,
$\ww_{  jklmnp}=C_{\aa jklmnp}$, the relevant couplings are
$\ww_2 DX F^3$ or $\ww _6 DX F$. Finally, the {\bf 8} could arise from
$C_{ijklmnp} $, $G_{\mm j}$ or $ B_{\mm j}$. In the first case, the coupling
would be
$C_7 \lll F$, while for the remaining two the linearised form of the coupling
to the vector $A_i$ would be
of the form
$A \lll DX \lll *F$.
The fact that M-theory compactified on $T^8$ gives {\bf 35+1+56} scalars and
{\bf 28+8  }
vectors (before duality transformations) suggests the
coupling to the {\bf 35+1+56} scalars
$G_{ij}, C_{ijklm}$ together with the
{\bf 28  }
vectors $\ww_2$ or $\ww_6$ and the
{\bf  8  }
vectors  $G_{\mm j}$ or $ B_{\mm j}$.
There are four choices of such action, one of which has the linearised form
$$
S= \int {1\over g^2} F\lll *F + C_5\lll F\lll F +\ww _6 \lll  DX\lll  F + A_1
\lll DX \lll *F
\eqn\ogg$$
(another is  the action obtained from this by replacing
 $\ww _6 DX F$ with $\ww_2 DX F^3$.)
It is intriguing that the action \ogg\ could be obtained by reducing the
10-dimensional action
$$
S= \int {1\over g^2} F\lll *F + C_6\lll F\lll F
\eqn\abc$$
on $S^1$, so that
the 9-dimensional theory on $T^8$ arises from the 10-dimensional theory on
$T^9$, and the $SL(9)$ subgroup of
$E_8$ arises geometrically.
The YM instantons give rise to 5-branes coupling to $C_6$, which reduce to give
  4- and 5-branes
in 8+1 dimensions.

\chapter {New Instantons and Branes}

In $6+1$ dimensional SYM, we have been led to consider the
coupling
$$
\int
\chi \Str DX^3 \lll F^2
\eqn\coup$$
This has many similarities with the term $\th F^2$ in $3+1$ dimensions.
First, we shall consider  solutions of the Euclidean version of 7-dimensional
SYM, whose bosonic
sector consists of 3 adjoint-valued scalars $X^\aa$ ($\aa=1,2,3$)
coupled to Yang-Mills, with the usual potential $\tr ([X^\aa, X^\bb])^2$.
Consider the following solution in $\R^7$. Splitting the coordinates $x^m$ into
coordinates $x^\aa$ on $\R^3$ and $x^i$ on $\R^4$, we choose $A_\aa =0$ and
$A_i$ the connection
for an instanton solution on $\R^4$, and choose
$X^\aa=
x^\aa t^\aa$
with no sum over $\aa$, where $t^\aa$ are any 3 mutually commuting
Lie algebra generators (not necessarily distinct) with product $t^1t^2t^3=T$.
Then the coupling \coup\ becomes
$$
\int
\chi \eta \lll
\tr  (T F\lll F)
\eqn\coupo$$
where $\eta=\6
\ee_{\aa \bb \gg}dx ^\aa \lll dx^\bb \lll dx^\gg$
is the volume form on $\R^3$.
For example, taking a gauge group $G=U(N)$, we can take
an $SU(N)$ instanton on $\R^4$ and take $t^1=t^2=t^3=T$ all as the $U(1)$
generator, so that
\coupo\ reduces to
$
\int
\chi \eta \lll \tr    F^2$ and, for constant $\chi$, is proportional to the
second Chern class of
the instanton.
Then the coupling constant $\chi$ is a $\th$-angle term corresponding to these
Euclidean solutions. Note that we could continue this solution back to a
Minkowski space solution
with one of the scalars $X^\aa$ depending linearly on time.
This solution can be thought of as a brane with 3 Euclidean dimensions, located
at a point in
$\R^4$.

The term \coup\ comes from the dimensional reduction of the term
$$\int \chi \tr F^5
\eqn\ten$$
 in ten dimensions.
Any YM configuration on $T^3\times M_7$ (where $M_7$ is a 7-manifold) with
non-zero 5-th Chern class, $\int \tr F^5 \ne 0$, and which is independent of
the toroidal
directions can be reduced to a configuration for which the term \coup\ is
non-zero.
For example, the configuration
considered above on $\R^4\times T^3$ (with the $x^\aa$ directions compactified)
arises from the ten-dimensional configuration
on $\R^4 \times T^6$ with an $SU(N)$ instanton on
$\R^4$ and a magnetic monopole on each of the three $T^2$ i.e. on each
$T^2$ there is a configuration with non-zero first Chern class, $\int \tr F \ne
0$.

The dimensional reduction of
\ten\ gives the term
$$
\chi
DX^n F^{5-n}
\eqn\abc$$
in $10-n$ dimensions, coupling to all $n$ scalars in the SYM multiplet in
$10-n$ dimensions.
This emerges from the D-brane action from the coupling to
an $n$-form $C_{\mm_1 ...\mm _n}$.
SYM in 10-n dimensions on  $\R \times T^{9-n}$ is related to M-theory
compactified to $n+2$ dimensions,
  and an $n$-form gauge field is dual to a scalar in $n+2$ dimensions.
Such a coupling could have arisen for   $d=9-n$ with $n=1,2,3,4,5$, but it was
only for
$d=6$ that the group theory demanded a singlet of $SL(d)$, and in that case it
was the only
candidate.

For the case $d=7$, two possible couplings of the SYM to a 6-form were
proposed, and in each case
the 6-form couples to a 5-brane inside the 7-brane. For the coupling $C_6 \lll
F$, the 5-brane
 arises
from a magnetic monopole configuration, as in [\Doug],  in which $\int F$ is
non-zero over the $T^2$ transverse to the 5-brane.  For the coupling $\ww _6
\lll DX \lll DX$, the
5-brane emerges from   configurations  in which $DX\lll DX $ is the volume form
for  the $T^2$ transverse to the 5-brane, e.g. with $X^\aa$ taking values in
the $U(1) $ subgroup of
$U(N)$ (generated by $T$) and identified with the coordinates on $T^2$, $X^\aa
=x^\aa T$.

\chapter{Conclusions}

We have used U-duality to obtain the couplings of the  SYM low-energy effective
action of the matrix
theory to the moduli for M-theory compactified on $T^d$, and considered some of
the BPS branes that
occur. For $d>3$, there are wrapped BPS $d-4$ branes corresponding to YM
instantons, and the
U-duality symmetry  mixes these with momentum modes. For $d=4$, the 0-branes
constitute an extra component of momentum and there is an extra hidden
dimension that becomes
manifest at finite coupling; the U-duality results from diffeomorphisms in this
higher dimensional
($5+1$) space.
For $d=5$, the U-duality $SO(5,5;\Z)$ is
a T-duality for the 5+1 dimensional string theory, relating
  string winding modes and momentum modes.
For $d>5$, there are $d-4$ branes that reduce to the $d=5$ strings in a
suitable limit, and so
the U-duality must relate momentum modes and $d-4$ brane wrapping modes.
However, there are
other backgrounds fields besides the torus metric $G_{ij}$ and the $d-3$ form
$C$ which couples to
the $d-4$ branes, and all of these are mixed by U-duality. As a result, the
U-duality mixes the momentum modes, the $d-4$ brane  wrapping modes and the
modes coupling to these
extra moduli, which are extra brane wrapping modes.

For $d=6$, there is one extra scalar modulus $\chi$, which couples to
instantonic branes, which
should play a similar role here to the instantons in 4-dimensional SYM; in both
cases, there is an
$SL(2,\Z)$ acting on the coupling constant and a $\th$-angle, and which can  be
thought of formally
as relating momentum modes and \lq $-1$-branes'. The $E_6(\Z)$ is then realised
as this
$SL(2,\Z)$ together with the $SL(6,\Z)$ acting geometrically on the 6-torus,
supplemented by
transformations relating   momentum modes (and \lq $-1$-branes') to membrane
wrapping modes.

For $d=7$, there are an extra 7 moduli corresponding to a vector or 6-form on
$T^7$, and hence to a
   vector or 6-form on $T^7\times \R$, and these should couple to an extra
0-brane or 5-brane.
If the coupling were to an extra 0-brane, this would combine with the
momentum modes in the {\bf 7} of $SL(7)$ to form
an {\bf 8} of the $SL(8)$ subgroup of $E_7$, which would imply the existence of
a hidden dimension
and a formulation in 8+1 dimensions on $T^8\times \R$.
However, it would be hard to write down a local 8+1 dimensional theory, as it
would have to couple to
a metric and a self-dual 4-form on $T^8$, and it is not clear how
to obtain such a 4-form from a field on $T^8\times \R$.
In fact, the D7-brane does not appear to couple to a 1-form, but does couple to
a 6-form,
corresponding to the fact that there are bound states of 5-branes inside
7-branes, but there
appear not to be ones of 0-branes inside 7-branes (there is no vector in the
type IIB theory and so
no conventional 0-branes).
 In this case, it appears that the theory should be in 7+1 dimensions, with
the $E_7(\Z)$  mixing momentum modes with 3-brane and 5-brane wrapping modes.

For $d=8$, there are an extra {\bf 28+8} moduli, corresponding to
an extra vector or 7-form and an extra 2-form or 6-form on $T^8$.
These would couple to an extra 0-brane or 6-brane and an extra 1-brane or
5-brane
in $T^8\times \R$. If the correct set of extra branes is a 0-brane plus a
5-brane, then
the
0-brane  combines with the
momentum modes in the {\bf 8} of $SL(8)$ to form
a {\bf 9} of the $SL(9)$ subgroup of $E_8$, which would imply the existence of
a hidden dimension
and a formulation in 9+1 dimensions on $T^9\times \R$.
The 5-brane and 4-brane in 8+1 dimensions would then arise from a
5-brane in 9+1 dimensions, and the $E_8(\Z)$ would act as the geometrical
$SL(9,\Z)$ together with
transformations mixing momentum modes with 5-brane wrapping modes.
Moreover, the fact that there is $SL(9)$, not $GL(9)$, suggests that this
should be a
scale-invariant theory in 9+1 dimensions.
Such a theory would presumably be invariant under the 10-dimensional
superconformal group discussed
in [\tonn].

For $d=9$, the counting is more subtle, but one might expect a theory on
$T^9\times \R$ with at least  a metric and 6-form on $T^9$, so that there is an
extra scale in the
theory compared with the $d=8$ case, so that this is presumably not
scale-invariant.
Indeed, the naive counting of scalar moduli is the same as that given in the
discussion of the
$d=8$ case in section 3, and gives 129 moduli, which indeed corresponds to
those of
$E_8/SO(16)$, plus an extra scale.
(The possibility that the extra modulus corresponds to the $\th$ angle in the
coupling $\th F^5$
would be at variance with the situation for lower $d$, where the scale always
enters in the
low-energy SYM theory.)

This suggests that the situation for $d=8,9$ is similar to that for $d=4,5$.
The D4 and NS5 brane both emerge from the M5-brane, while it is conjectured
that the D8 and NS9 brane
both emerge from an M9-brane [\GravDu]. This suggests that the required 9+1
dimensional theories both
emerge from the world-volume theories of wrapped M9-branes. A candidate for a
suitable theory has
been proposed in [\mar]; it is the target space theory of the (2,1) string.

\ack
{I would like to thank Mohab Abou Zeid, Bobby Acharya, Mike Douglas, Jos\' e
Figueroa-O'Farrill, Jerome Gauntlett and
Bernard Julia  for useful discussions.}

\refout
\bye